\documentclass[number,preprint]{elsarticle}
\usepackage[utf8]{inputenc}
\usepackage[T1]{fontenc}
\usepackage{wrapfig}
\usepackage{float}
\usepackage{flafter}
\usepackage[usenames, dvipsnames]{xcolor}
\usepackage{amsmath}
\usepackage{amssymb}
\usepackage{mathrsfs}
\usepackage{amsfonts}
\usepackage{makeidx}
\usepackage{enumitem}
\usepackage{tabularx}
\usepackage{stmaryrd}
\usepackage{mathtools}
\usepackage{soul}
\usepackage{bm}
\usepackage[left=2.5cm,right=2.5cm,bottom=2.5cm,top=2.5cm]{geometry}
\usepackage{graphicx,subcaption}
\usepackage{hyperref}
\newcommand{\vect}{\mathbf}						    
\newcommand{\etimes}[1]{\times 10^{#1}}             

\newcommand{\vv}{\vect{v}}							
\newcommand{\of}{\Omega_{\mathrm{f}}}

\newcommand{\ii}{\mathrm{i}\mkern1mu}		         
\newcommand{\dd}{\mathrm{d}}						 

\title{Enhancement of the sound absorption of closed-cell mineral foams by perforations: Manufacturing process and model-supported adaptation}
\author[1]{Bart Van Damme\corref{cor1}}
\ead{bart.vandamme@empa.ch}
\author[1]{Théo Cavalieri\fnref{fn1}}
\author[2]{Cong-Truc Nguyen}
\author[2]{Camille Perrot}
\cortext[cor1]{Corresponding author}
\fntext[fn1]{Now at: Laboratoire d'Acoustique de l'Université du Mans (LAUM),
UMR 6613, Institut d'Acoustique-Graduate School (IA-GS), CNRS,
Le Mans Université, Avenue Olivier Messiaen, 72085 Le Mans, France}
\affiliation[1]{organization={Empa, Materials Science and Technology},
addressline={Ueberlandstrasse 129},
postcode={8600},
city={Dübendorf},
country={Switzerland}}
\affiliation[2]{organization={Univ Gustave Eiffel, Univ Paris Est Créteil, CNRS, UMR 8208, MSME},
postcode={F-77454},
city={Marne-la-Vall\'ee},
country={France}}
\date{\today}
\begin{document}

\begin{abstract}
Thin low-frequency acoustic absorbers that are economical to produce in large quantities are scarce, and their efficiency is often limited to a narrow frequency range. In this paper, we present opportunities to use highly porous mineral foams, in particular optimally designed gypsum foams, to achieve high absorption levels for layers of less than 1/10 of a wavelength thick. To reach this goal, we perforate a fraction of the initially closed pores using thin needles. Finite element simulations of the fluid flow in a representative volume element show how the combination of foam properties (cell size and wall thickness) and perforation pattern (hole diameter and perforation distance) can be chosen such that sub-wavelength absorption is obtained. In particular two transport parameters used in the approximate but robust Johnson-Champoux-Allard model for porous media have to be optimized: the flow resistivity and high-frequency tortuosity. The fluid flow modeling results are successfully compared with sound absorption measurements, showing indeed that the proposed material, once appropriately perforated, yields a remarkable low-frequency sound absorption peak. On a more fundamental level, this paper shows how the multiporosity, the presence of microcracks, and the material's surface roughness can be exploited to enhance its acoustic absorption at very low frequencies.
\end{abstract}
\maketitle
\section{Introduction}
\label{sec:Introduction}
Sound absorbing materials are key to creating a pleasant acoustic environment, both in- and outdoors.
Reducing the reflected acoustic energy lowers the overall sound pressure level (SPL), increases speech intelligibility, or enhances musical performance and recording quality. Porous absorbers are ubiquitous since they are affordable and easy to manufacture, but they require thick layers to be effective at low frequencies~\cite{moeser2009book,Allard2009}.
The thickness of an absorbing material is typically expressed as a fraction of the corresponding wavelength of the sound frequencies that are well absorbed. In the case of classical porous media, at least a quarter wavelength is required to assure effective absorption.
This is practically problematic for the frequency range below 500~Hz, where a large surface covered by at least 0.2~m absorber material is needed. 

Several space-saving solutions for low-frequency absorption do exist. Plate or membrane absorbers trap sound wave energy in structural vibrations of a thin plate, tuned to a specific resonance frequency~\cite{frommhold1994membrane,moeser2009book}. Perforated plates backed by an air-gap show a prominent, but narrow, low-frequency absorption peak~\cite{maa1998perforated,atalla2007perforated}. Recently, the field of acoustic metamaterials has opened ways to achieve absorption profiles `on demand', using well-designed structures including membranes~\cite{yang2008membraneMM,mallejac2019zero} or Helmholtz resonators~\cite{lee2009acousticMM,haberman2016acoustic,jimenez2021acoustic}. The absorption bandwidth of these solutions is typically narrow, but can be broadened and shifted by either adding classical foam absorbers~\cite{atalla2007perforated,li2021perfporor}, or using multiple resonating structures~\cite{jimenez2017rainbow,liu2018rainbow}.

The acoustic properties of porous media can be efficiently modelled using a homogenization approach, where an effective, frequency-dependent and complex valued bulk modulus and mass density are attributed to the material~\cite{Allard2009}. 
The semi-phenomenological Johnson-Champoux-Allard (JCA) model makes it possible to relate the foam's macroscopic properties, in particular its transport parameters which could be derived from geometry and the interaction with the saturating fluid, to homogenized material parameters~\cite{johnson1987theory,champoux1991dynamic}. Five so-called transport parameters, i.e. static airflow resistivity, dynamic tortuosity, porosity, and thermal and viscous characteristic length, describe the material sufficiently for commonly encountered porous materials to calculate its homogenized mass density and bulk modulus. The five parameters can be retrieved experimentally or via pore-scale simulations. 
The latter micro-macro approach has already proven to be effective in the case of open-cell foams~\cite{perrot2008bottomup,perrot2012microstructure,chevillotte2017threedim}, fibrous materials~\cite{mamtaz2017acoustic,luu2017fibrous}, perforated plates~\cite{atalla2007perforated}, and perforated foams~\cite{chevillotte2010perforated}.

For many applications, it is desirable to achieve an absorption spectrum by design, in order to resolve acoustical issues within a certain frequency range. 
Tuning the membrane content could be used as a means to control sound absorption at constant pore size~\cite{trinh2019tuning}. Additive manufacturing techniques were also employed in order to design sound absorbers based on periodic open cells~\cite{DESHMUKH2019107830,WANG2023112130,boulvert_optimally_2019}, but the acoustic properties depend a lot on the printing process~\cite{zielinski_reproducibility_2020}. It is also known that perforating open- or closed-cell foams can enhance their absorption performance, or shift the peak absorption frequency~\cite{sgard2005perforated,chevillotte2010perforated,opiela2021perforated}. Both large holes and sub-millimeter (or micro-) perforations through the thickness of the porous layer can lead to an improved absorption level, based on different physical principles: 
Perforating initially closed-pore materials leads to partially open porosity, and the acoustic properties are purely determined by the perforations~\cite{opiela2021perforated}. 
In the case of adding perforations to media with initial open porosity, pressure diffusion between phases with high and low flow resistivity can enhance the absorption properties~\cite{olny2003double,sgard2005perforated}. Finally, adding perforations can induce an additional scale of open pores, leading to double-porous materials~\cite{chevillotte2010perforated}. 
An overview of the effect of perforating microporous media was given by \cite{sgard2005perforated}, who pointed out that improving the combined absorption properties typically benefits from large holes placed far apart. 

Even though perforations can alter the acoustic properties of a porous material, achieving a desired absorption curve is highly limited by the base material. 
In this paper, we investigate the innovative approach of adding micro-perforations to an interesting class of rigid, low-densty, mineral foams to achieve low-frequency acoustic absorption, even in the presence of a rigid backing. 
The goal of our proposed material is to reach an absorption peak for layers thinner than one tenth of a wavelength, which is often referred to as sub-wavelength sound absorbers. 
The proposed mineral foams, made out of gypsum, cement, or even ceramic materials, show excellent thermal insulation properties, are non-flammable, and exhaust no toxic particles. 
Thanks to the specific manufacturing process, a total porosity of up to 95\% can be achieved, and the mass density can be below $100\,\mathrm{kg.m^{-3}}$. 
However, after production the pores are almost entirely closed, leading to a poor acoustic absorption performance since air particles cannot penetrate the material. 
The pores can be opened by perforating the foams, thereby creating window-like connections between the pores. 
This leads to a double-porous material, with micro-pores defined by mineral crystals in the micrometer-range, and (meso-)pores defined by the foaming process in the millimeter-range. 
It is well known that multiporous materials have advantageous acoustic properties~\cite{olny2003double,gourdon2010double,becot2008double,venegas2023multi}. 
This work will show that a well-chosen perforation size and pattern, combined with optimal pore properties, can alter the acoustic absorption curve to the desired frequency range.
In terms of transport parameters, the pore size, wall thickness, and perforation properties give rise to an exotic combination of the flow resistivity, tortuosity, and viscous characteristic length. 
The absorption is further enhanced by the solid frame's microporosity, which depends essentially on the chemical components of the mineral suspension. 

The article is organized as follows: first, the foam manufacturing process and the acoustic homogenization problem for this particular material are described in Sec.~\ref{sec:Problem}, focusing on the complexity of defining a representative volume element.
Then a parametric study of the adjustable geometric properties, namely the pores and perforations, is conducted and presented in Sec.~\ref{sec:Parametric}.
The experimental studies to validate the model are given in Sec.~\ref{sec:Experimental}, and the conclusions of this work are presented in Sec.~\ref{sec:Conclusions}.
\section{Homogenization problem for highly porous perforated foams}
\label{sec:Problem}
\subsection{Manufacturing of low-density mineral foams with closed pores}\label{sec:Manufacturing}
The gypsum foams are prepared from calcium sulphate hemihydrate, mixing water, a foam formation powder (FFP), a blowing agent, and a catalyst to trigger the blowing reaction. In a first step, foam formation powder and catalyst are well dispersed in the mixing water. Then, the calcium sulphate is admixed, and the suspension stirred for 5 min. Subsequently, the blowing agent is added and well homogenized for 10 s whereupon the mixture is poured into a mold where it is given time to fully expand. After a few minutes, when a strength gain is manually sensed, the foam is demolded, cut to the desired dimensions, and optionally perforated. In a last step, the material is dried at temperatures below 40°C.
The dry foam density $df$ is determined by dividing the weight of a cuboid sample by the volume of the cuboid. The total porosity is calculated as $\phi_t = 100 \times (1-df/dg)\%$, where $df$ denotes the foam density and $dg$ the theoretical density of gypsum. The pore size distribution is obtained by analyzing cross-sectional optical images with the software Lince (Linear Intercept, TU Darmstadt, Germany).

The FFP~\cite{deCavis2016Patent} plays a key role in the process. It comprises partially hydrophobized particles~\cite{gonzenbach2006stabilization} that form a particle shell~\cite{sturzenegger2014particle} at the air-water interface during expansion and yield wet foams of unprecedented stability. Particle-stabilization is a prerequisite to create gradient-free, homogeneous foams with tailored pore size.
The porosity is empirically determined by the blowing agent concentration, while the pore size can be adjusted by varying the amount of FFP and blowing agent. It decreases with increasing FFP addition. In case the desired pore size cannot be achieved varying the amount of FFP alone, the blowing agent concentration is varied in two experiments by $\pm$20\% and the median pore size plotted as a function of peroxide concentration. The correlation between pore size and peroxide concentration is linear in a first approximation.

The microporosity of the gypsum foam can be manipulated by changing the water to solids ratio w/s. An increase in w/s enlarges the micropore fraction, a decrease lowers it. The w/s minimum is given if the slurry is too viscous for processing, whereas the maximum w/s is limited by the minimal mechanical strength requirements for the foam. The geometry of the obtained gypsum foams will be described in the next section. The reliability and reproducibility of the foam manufacturing process in terms of pore size (and its distribution), porosity, and density, will be illustrated.

\subsection{Pore structure and perforation geometry}
\label{sec:Geometry}
\begin{figure}[H]
\centering
\includegraphics[scale=1]{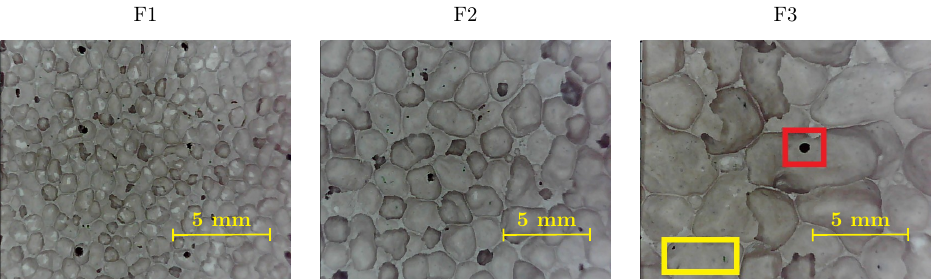}
\caption{(Color online) Optical microscope image of the outer surface of three investigated gypsum foams with increasing pore size. A perforation with diameter 0.5~mm is highlighted in red.
All samples are generally closed-pore materials but show small cracks in several cell walls, e.g. in the area highlighted in yellow.}
\label{fig:ThreeSamples}
\end{figure}

Gypsum foams are well known to exhibit combined mechanical and thermal properties~\cite{Capasso2021,Liu2023}.
In this paper, we investigate the effect of perforations in rigid gypsum foams to enhance their acoustic absorption properties. Here, we focus on foams with a pore size ranging from 1 to 3~mm, a total porosity over 89\%, and a wall thickness around 0.1~mm.
Optical microscope images of three different materials with increasing pore size are shown in Fig.~\ref{fig:ThreeSamples}, and their physical properties (pore size, total porosity $\phi_\mathrm{t}$, and mass density) are listed in Tab.~\ref{tab:threesamples}. Their values were retrieved from 100~mm$\times$100~mm$\times$20~mm samples, thus including a large amount of pores for the analysis.
\begin{table}[H]
    \centering
    \caption{Physical properties of three samples of gypsum foam}
    \begin{tabular}{c|c|c|c}
      material   &  median pore size $2R_p$ (mm) & total porosity $\phi_\mathrm{t}$ (vol\%) & density $\mathrm{(kg.m^{-3})}$\\
      \hline
       F1  & 1.03 & 89.7 & 238 \\
       F2  & 2.04 & 92.1 & 182 \\
       F3  & 3.28 & 94.6 & 124
    \end{tabular}
    
    \label{tab:threesamples}
\end{table}
The manufacturing process is extremely reproducible: Based on an analysis of 17 large samples (420~mm$ \times$ 420~mm $\times$ 400~mm) of material F3, the standard deviation of the density was only 2.3\% and of the median pore size 3.3\% of their respective average values. The materials are very homogeneous, showing less than 1\% deviation in the values for samples taken from the top and bottom of the cast foam blocks. The span of the pore size is defined as $S_p = (d_{90}-d_{10})/d_{50}$, with $d_N$ being the $N$-percentile pore diameter. This inherent variability for a given material was analyzed for material F3, showing $S_p=1.16$~mm. 

The pores are opened by an array of perforations with a diameter larger than inter-crystalline porosity but significantly smaller than the pore size. Assuming an initial perfectly closed-cell material with rigid walls, the array of opened pores governs the permeability to acoustic waves, and therefore the material's acoustic parameters. If the spacing is larger than the average pore size, not every pore will be opened, leading to an open porosity $\phi<\phi_\mathrm{t}$. The other transport parameters also depend on the perforation properties. It is clear that the perforation diameter and the number of perforations per unit area control the flow resistance. A larger distance between perforations increases the tortuosity due to the flow distortion when entering the material, as shown by~\cite{atalla2007perforated}. A detailed study of how the transport parameters are controlled by the geometry is given in sec.~\ref{sec:Parametric}. 

Having the possibility to modify the transport parameters by the perforation properties allows us to tune the absorption spectrum for a given foam and prescribed material thickness. In what follows, we investigate a square pattern of perforations with a spacing between two perforations ranging between 3 and 10~mm. A spacing smaller than the pore size would open all pores, leading to the equivalent of an open-pore material as we will show in the following sections. Similar to typical dimensions for micro-perforated plates~\cite{maa1998perforated}, we choose the perforation diameter between 0.3 and 0.8~mm. This is small enough to create window-like connections in faces between individual pores without altering their shape and structure too much.

\begin{figure}[h]
\centering
\includegraphics[scale=1]{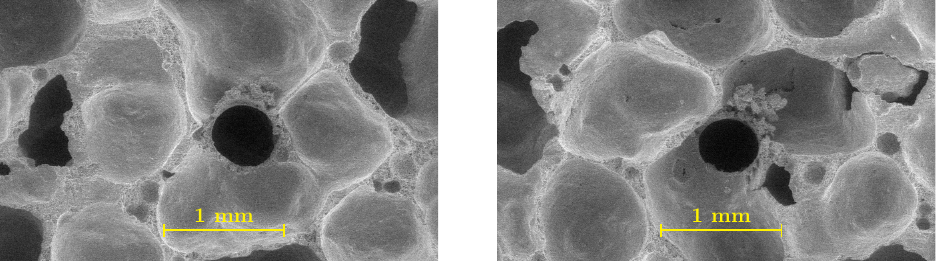}
\caption{Scanning electron microscope (SEM) image of two perforations with diameter 0.5~mm in the F3 foam sample, showing the roundness of the holes and their possible positions with respect to the pores.}
\label{fig:Perforation}
\end{figure}

\begin{figure}[H]
\centering
\includegraphics[scale=1]{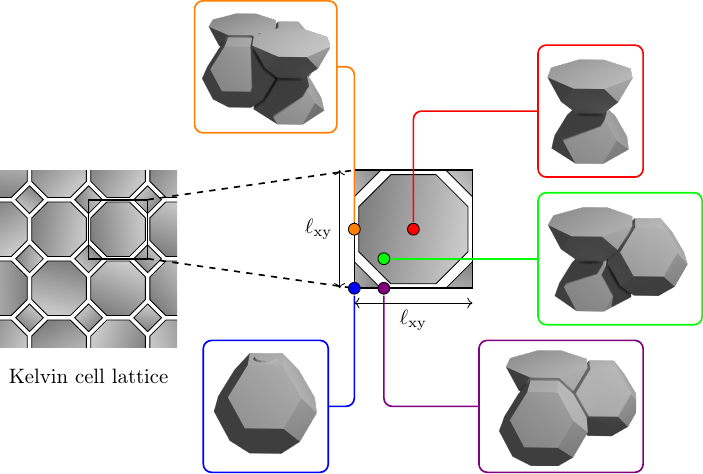}
\caption{(Color online) Possible arrangements of connected pores depending on the perforation position in the Kelvin cell lattice. The Kelvin cell lattice is shown on the left, the three-dimensional pore arrangements are shown on the right, together with the corresponding RVE.
Open-pore structure with perforation position $\mathbf{p}=(0,0)$ in blue, $\mathbf{p}=(\ell_\mathrm{xy}/4,0)$ in violet, $\mathbf{p}=(\ell_\mathrm{xy}/4,\ell_\mathrm{xy}/4)$ in green, $\mathbf{p}=(0,\ell_\mathrm{xy}/2)$ in orange, and $\mathbf{p}=(\ell_\mathrm{xy}/2,\ell_\mathrm{xy}/2)$ in red.}
\label{fig:Lattice}
\end{figure}

\begin{figure}[h]
\centering
\includegraphics[scale=1]{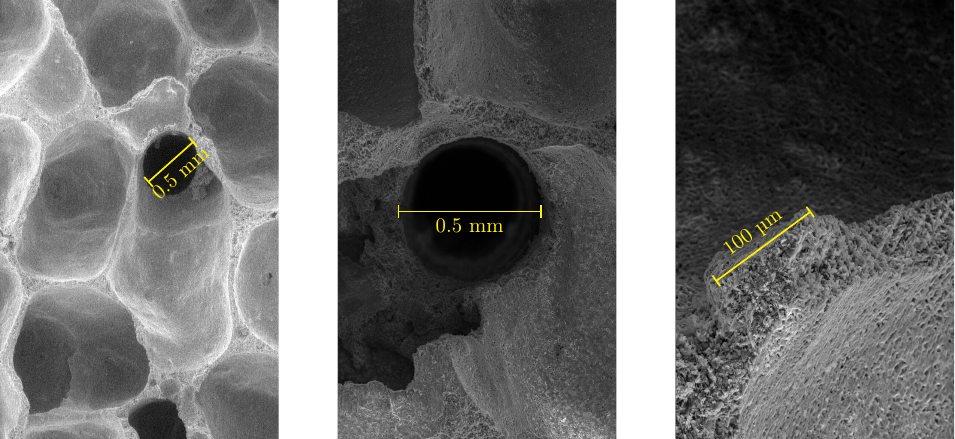}
\caption{SEM images in the vicinity of a perforation, illustrating different features and scales of the perforated gypsum foam: pore size (left panel), perforation size (central panel), gypsum matrix (right panel).}
\label{fig:Zoom}
\end{figure}

An illustration of some typical perforations of a gypsum foam sample is provided in Fig.~\ref{fig:Perforation}. It can be seen that perforations can occur in a face perpendicular to the needle, thereby connecting two pores as illustrated in the schematic of (Fig.~\ref{fig:Lattice}, blue or red configurations); but the needles can also go through a parallel face, connecting more than 2 adjoining pores (Fig.~\ref{fig:Lattice}, other configurations). Perforating the samples with needles shortly after the foaming process ends, i.e. when the cell walls haven't completely hardened out yet, ensures that the holes are round and the cell walls are not fractured. Therefore, we can assume that all perforations are identical with a radius equal to that of the perforating tool.
It is well known that gypsum is by itself a porous material~\cite{chevillotte2013doublepor}, with micropores between the gypsum crystals that are much smaller than the cells created by the foaming process. The microporosity of the cell walls can clearly be seen in Fig.~\ref{fig:Zoom}. In between the gypsum crystals, air pockets with a size below 10~$\mu$m are clearly visible. The co-existence of pore sizes differing by several orders of magnitude requires special care to model the homogenized acoustic properties~\cite{olny2003double,chevillotte2013doublepor}.
\subsection{Defining a Representative Volume Element}
\label{sec:RED}
Deriving homogenized fluid properties of complex porous media is well described in literature, and the main points are recapitulated in \ref{sec:Homogenization}. 
The JCA homogenization approach requires the knowledge of five so-called transport parameters: open porosity, static air flow resisitivity, dynamic tortuosity, and the thermal and viscous characteristic lengths.
The calculation of the five transport parameters requires a numerical simulation of viscous and inviscid flows through the porous material.
To allow a parametric study on the effect of the perforations without excessive computational time, the modelling domain should be kept as small as possible. 
A Representative Volume Element (RVE) should contain enough information such that the transport parameters and the effective properties can be calculated on the sole basis of its volume. 
In open-cell porous foams with inter-connected pores, reducing the modelled volume is typically done by defining a unit-cell which can be periodically duplicated by translation in all directions and is therefore called a periodic unit cell (PUC).
The classical model for isotropic foams uses a regular tetrakaidecahedron (Kelvin cell) as the PUC~\cite{perrot2012microstructure}.
It is known however that a distribution of pore sizes and wall thickness can enhance the acoustic properties, and it can be included in the models~\cite{nguyen2024effect}.
In the case at hand, the perforation pattern does not coincide with the periodicity of the Kelvin cell, and therefore a PUC cannot be employed.
Depending on the location of the perforations $\mathbf{p}$ within a regular arrangement of Kelvin cells with lattice constant $\ell_\mathrm{xy}$, the connected cells within the RVE ranges from two to six connected (parts of) pores (Fig.~\ref{fig:Lattice}). 

Because the perforations and the Kelvin cell lattice do not have the same periodicity, a PUC cannot be defined. We therefore suggest a slightly modified micro-macro approach \cite{auriault_homogenization_2009,sanchez-palencia_non-homogeneous_1980,bensoussan_asymptotic_1978}, taking  the interconnected pore pattern resulting from the perforation of closed pores into account as the main feature defining the acoustic properties.

\begin{figure}[h]
\centering
\includegraphics[scale=1]{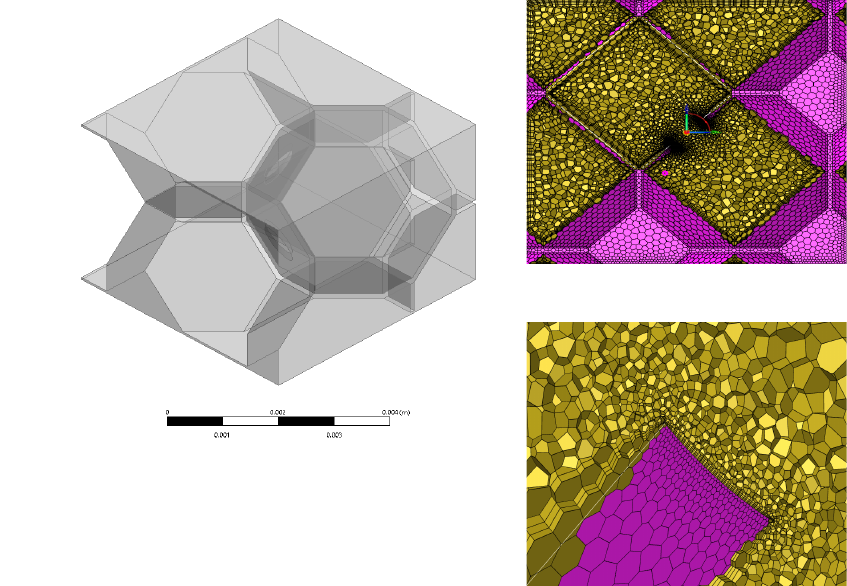}
\caption{(Color online) RVE geometry (left) and cross-section of the mesh in Ansys Fluent (right) showing a detail of the mesh refinement at the perforation edge.
Volume elements are shown in yellow, the face discretization in purple.}
\label{fig:Mesh}
\end{figure}

We propose a RVE consisting of a cut-out of an assembly of Kelvin cells, with lateral dimensions equal to the perforation spacing, and centered around the perforation. Given the relative area of the inclined faces within the Kelvin cell, the perforation highlighted in green in Fig.~\ref{fig:Lattice} is most likely to occur, and the three-pore combination will be used for the numerical implementation.
Through the thickness, a single Kelvin cell is chosen. 
A thin inlet and outlet are added at the top and bottom of the RVE to allow a fluid to make contact with all pores exposed to the impinging sound wave.
This choice is supported by the reasoning that air particles enter a perforation only when they are within the area of the RVE.  
They therefore always enter the foam through the closest perforation, regardless of the exposed surface geometry.
Although the velocities in the perforation are high compared to the rest of the RVE, the formulas require averaging/integration over the entire volume and surface. 
Therefore, all pores facing the impinging sound wave have to be taken into account.
The important role of the inlet layer is clarified in \ref{sec:inlet}.
This analysis shows that adding an inlet layer, thereby including the remaining closed pores into the model, mainly changes the value of the tortuosity. 
The flow lines are heavily distorted by the rugosity of the inlet surface when entering the perforation, and this gives rise to an increased value of the dynamic tortuosity. 
In the lateral directions, no interconnected pores exist and periodic boundaries are not necessary.
An example of the resulting RVE's geometry is shown in Fig.~\ref{fig:Mesh}, for a perforation diameter equal to 0.5~mm and a cell size of 3~mm. 
The agreement between models and experimental results will show that the assumptions leading to this RVE are astute. 
\subsection{Implementation of the homogenization models}
The viscous and inviscid models as described in ~\ref{sec:Homogenization} are implemented in the commercial software Ansys Fluent (version 2022 R2). The perforated Kelvin cell RVE is fully parameterized, so that the important geometric properties can be changed at will: pore radius $R_\mathrm{p}$ (defined here as the distance between the pore center and the hexagonal faces, which is smaller than the distance to the square faces), wall thickness $t$, perforation radius $r$, and perforation distance $d$. The perforation is centered on a hexagonal face, thereby connecting two externally exposed half-pores to an internal pore. This gives rise to geometries as shown in Fig.~\ref{fig:Mesh}, showing the interconnected pores and additional closed pores connected to the inlet layer. 

To allow for accurate microscopic inviscid velocity calculations on the face of the perforation (the Laplace equation,~\ref{sec:Homogenization}), the surface mesh is refined to a size of 5 \textmu m on the perforation edges.
In order to avoid numerical errors for the Stoke's flow, the mesh contains a three-element boundary layer on every surface.
The fluid is chosen to be air, with a mass density of $\rho_0=1.225\,\mathrm{kg.m^{-3}}$ and a dynamic viscosity $\mu = 1.89\etimes{-2}\,\mathrm{Pa.s}$.
The boundary conditions are as follows:
\begin{itemize}
\item The pressure inlet coinciding with the top surface is set as a Dirichlet boundary condition with $P=1\,\mathrm{mPa}$ in order to ensure laminar flow. The electrical potential of this face is $q=100\,\mathrm{V}$. 
\item The pressure outlet on the bottom face is set to $P=0\,\mathrm{Pa}$ and an electric potential of $q=0\,\mathrm{V}$.
\item All other faces have no-slip flow conditions $\vv\vert_{\partial\Omega} = \mathbf{0}$, with a no-penetration condition for electric current.
\end{itemize}
\section{Parametric study of the transport parameters}
\label{sec:Parametric}
In order to assess the variation of effective properties --the bulk modulus and mass density, resulting in the absorption coefficient -- with respect to the geometric features of the porous structure, a parametric study is conducted. On the one hand, we study the effect of pore size and wall thickness in Sec.~\ref{sec:Foam}, and on the other hand the dependence of the effective properties with respect to perforation is studied in Sec.~\ref{sec:Lattice}.

The standard configuration is based on one of the most promising combinations: material F3, which is a gypsum foam with a pore diameter of $2R_p = 3$~mm and wall thickness $t = 0.1$~mm, resulting in a total porosity of 94.6\%. The perforation diameter is $2r = 0.5$~mm and the lattice distance is $d=5$~mm. This combination leads to a surface perforation rate of 0.79\%. Assuming all pore walls are closed, except for the perforated walls, the geometric analysis shows a total open porosity of 70\%, meaning that more than a quarter of the pores is not connected due to perforations. The following parametric studies show the influence of changing one parameter while keeping the other ones identical to the state in the basic configuration.
\subsection{Foam properties: pore radius and wall thickness}
\label{sec:Foam}
The foam manufacturer is able to produce foams with median pore sizes up to 4~mm. Given the practical limitations regarding brittleness and mass density, we investigate pore sizes in the range lying between 1 and 3~mm corresponding to an acceptable trade-off between lightness and mechanical strength. The second foam property that can be changed is the wall thickness. For the given pore sizes, the manufacturing process allows this parameter to be changed from 60 to 200~\textmu m. 
Changing the amount of mixing water in the suspension controls the microporosity of the cell walls between 20\% and 40\%, 
which can be taken into account for the calculation of the total equivalent fluid properties~\cite{chevillotte2013doublepor}, but does not enter into the numerical simulations.
\begin{figure}[h]
\centering
\includegraphics[scale=1]{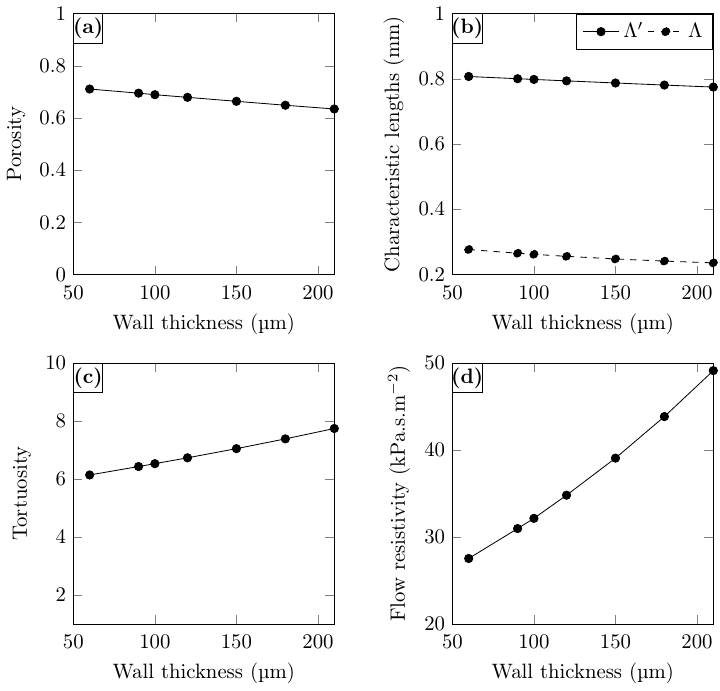}
\caption{Simulated transport parameters for varying  wall thickness, with $2R_p=3$~mm, $d=5$~mm, and $2r=0.5$~mm. The open porosity $\phi$ is shown in \textbf{(a)}, the high-frequency limit of tortuosity $\tau^\infty$ in \textbf{(b)}, the characteristic thermal and viscous length (respectively $\Lambda^\prime$ and $\Lambda$) in \textbf{(c)}, and the static air-flow resistivity $\sigma$ is shown in \textbf{(d)}.}
\label{fig:ParametricWallThickness}
\end{figure}
\begin{figure}[h]
\centering
\includegraphics[scale=1]{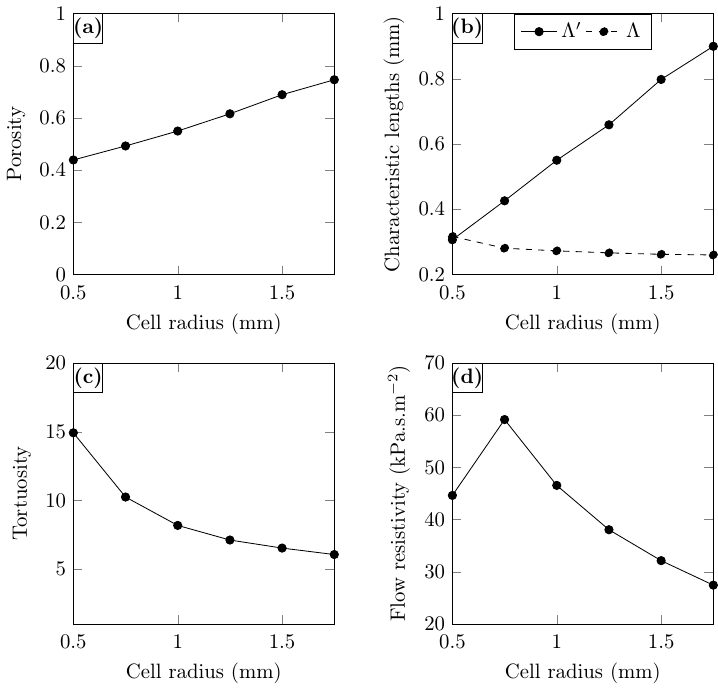}
\caption{Simulated transport parameters for varying cell radius, with $t = 0.1$~mm, $d=5$~mm, and $2r=0.5$~mm. The open porosity $\phi$ is shown in \textbf{(a)}, the high-frequency limit of tortuosity $\tau^\infty$ in \textbf{(b)}, the characteristic thermal and viscous length (respectively $\Lambda^\prime$ and $\Lambda$) in \textbf{(c)}, and the static air-flow resistivity $\sigma$ is shown in \textbf{(d)}. For small pore radii the perforations go through multiple faces at once, and change the monotonic variation of the flow resistivity. 
}
\label{fig:ParametricCellRadius}
\end{figure}
An overview of the computed transport parameters as a function of the wall thickness is shown in Fig.~\ref{fig:ParametricWallThickness}. This parameter has little influence on the open porosity, thermal, and viscous length. The tortuosity and static flow resistivity both increase with increasing wall thickness. By contrast, the cell radius has a more important effect on most transport parameters as shown in Fig.~\ref{fig:ParametricCellRadius}. Both the porosity and thermal length increase with increasing pore size. Contrary to $\Lambda^\prime$, $\Lambda$ is a dynamically connected pore radius influenced by the local field $\vect{E}$ which strongly increases in the vicinity of the perforation, Eq.~\eqref{eq:lbd}. Therefore, the thermal (resp. viscous) length is typically of the order of the pore (resp. perforation) radius and the viscous length, which is confirmed in our case. The tortuosity and flow resistivity both decrease with increasing pore size. 
For a pore radius smaller than 0.75~mm, the flow resistivity seems to drop again. In this case, the cell is so small that both diagonal and horizontal walls are perforated. The viscous flow has significantly different characteristics, breaking the monotonic increase of flow resistivity. Surprisingly, the equal viscous and characteristic length point to a tube-like, rather than an open-pore behaviour. The tortuosity however remains high thanks to the rough surface around the perforated cells.
For very small pores ($R\to 0.5$mm, $R>r$; Fig.~\ref{fig:ParametricCellRadius}c), extremely high values of the tortuosity can be achieved. However, this also yields a too high flow resistivity to create effective sound absorbers.
\subsection{Perforation properties: diameter and lattice distance}
\label{sec:Lattice}
The perforation pattern allows us to tune the acoustic properties of the foams if they are combined with a suitable base foam. As described in ~\ref{sec:Homogenization}, the perforations determine the open porosity. 
We assume that the perforation diameter is smaller than the wall size of a pore in order to maintain a typical foam structure with windows in individual cell walls. We therefore vary the possible perforation diameters between 0.2 and 1~mm. The distance between two perforations is also chosen to be at least the diameter of a pore, to avoid multiple perforations inside the same pore. We choose a range from 3~mm to 11~mm for the perforation distance. However, even for the smallest perforation distance the material is not fully equivalent to an open-cell material 
due to a lack of lateral connections. This leads to anisotropic acoustic properties, although the values of the transport parameters in the remaining principal directions are not further investigated in this paper. 
\begin{figure}[h]
\centering
\includegraphics[scale=1]{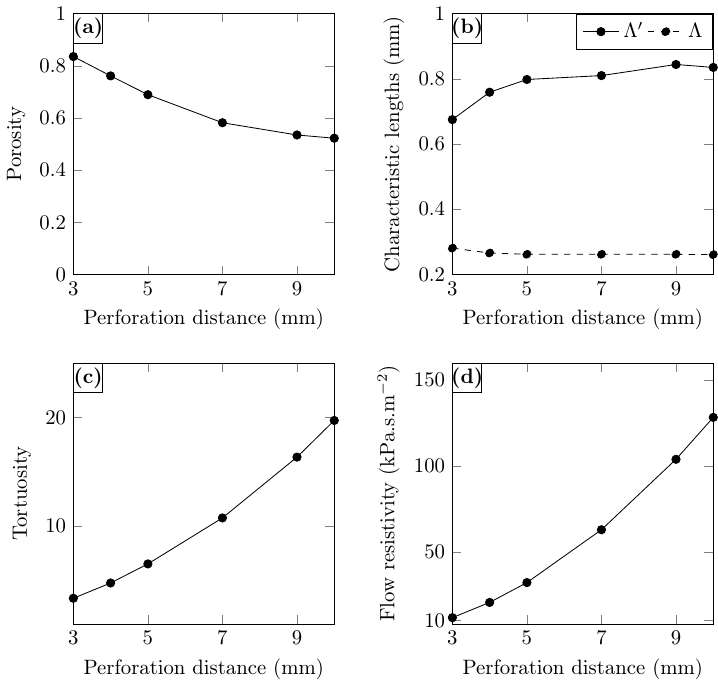}
\caption{Simulated transport parameters as a function of the perforation distance, with $t = 0.1$~mm, $2R_p=3$~mm, and $2r=0.5$~mm. The open porosity $\phi$ is shown in \textbf{(a)}, the high-frequency limit of tortuosity $\tau^\infty$ in \textbf{(b)}, the characteristic thermal and viscous length (respectively $\Lambda^\prime$ and $\Lambda$) in \textbf{(c)}, and the static air-flow resistivity $\sigma$ is shown in \textbf{(d)}. 
}
\label{fig:ParametricPerforationDistance}
\end{figure}
\begin{figure}[h]
\centering
\includegraphics[scale=1]{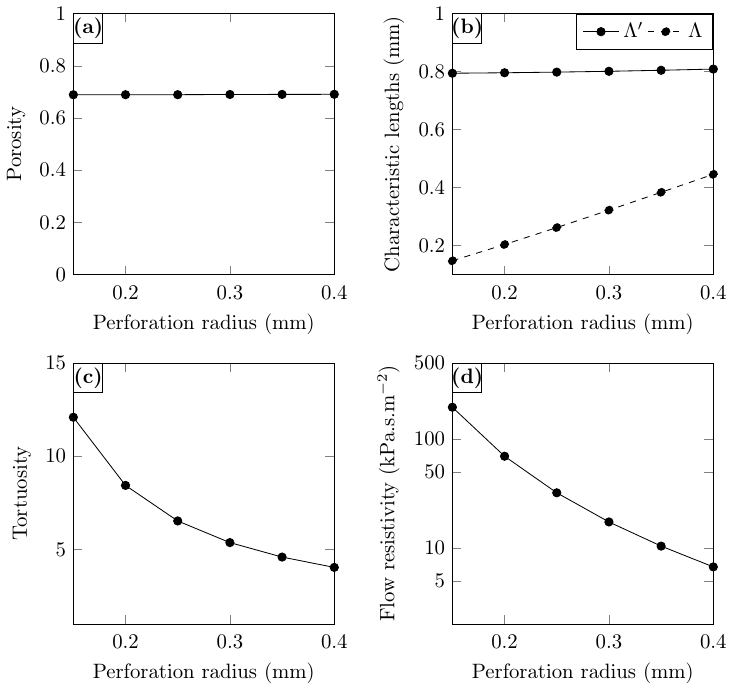}
\caption{Simulated transport parameters for varying perforation radius, with $t = 0.1$~mm, $2R_p=3$~mm, and $d=5$~mm. The open porosity $\phi$ is shown in \textbf{(a)}, the high-frequency limit of tortuosity $\tau^\infty$ in \textbf{(b)}, the characteristic thermal and viscous length (respectively $\Lambda^\prime$ and $\Lambda$) in \textbf{(c)}, and the static air-flow resistivity $\sigma$ is shown in \textbf{(d)}.}
\label{fig:ParametricPerforationDiameter}
\end{figure}
The effect of varying the perforation distance on the transport parameters is shown in Fig.~\ref{fig:ParametricPerforationDistance}. It is observed that the perforation distance has a very important influence on the tortuosity and flow resistivity, which both increase when the perforations are further apart. Further discussions of these results will be provided in the next section. Once again, although higher tortuosity might be desirable to achieve low-frequency sound absorption, the perforation pattern leads to an overly high flow resistivity and therefore reduced maximal sound absorption.
The lower open porosity also adversely affects the absorption properties.

In agreement with our previous analysis, the perforation radius governs the value of the viscous characteristic length, while not significantly changing the open porosity and thermal characteristic length (Fig.~\ref{fig:ParametricPerforationDiameter}). The tortuosity and flow resistivity decrease with increasing perforation sizes. Given the simulation results, varying the perforation diameter has a significantly larger effect on the flow resistivity than all other investigated absorber parameters.
\subsection{Discussion of the main results of the parametric study}
\begin{figure}[h]
\centering
\includegraphics[scale=1]{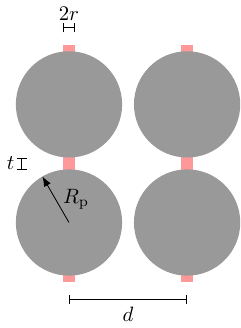}
\caption{(Color online) Sketch of the perforated cell geometry, showing the four design parameters. The pores are shown as circles in light blue, and the perforations are in green.}
\label{fig:UnitCell}
\end{figure}

The quantitative findings of the previous study can be mostly explained in an intuitive way. The static flow resistivity increases for smaller pores, thicker cell walls, smaller perforations and larger perforation distance, which can be explained by the schematic representation in Fig.~\ref{fig:UnitCell}. Since the flow resistance $R_f$ is mainly governed by the narrow cylindrical perforations, its value can be approximated by the analytical formula for cylindrical tubes as
\begin{equation}
    R_\mathrm{f} = \frac{\Delta p}{Q_\mathrm{vol}}
    \;,
    \label{eq:FlowResistance}
\end{equation}
where $\Delta p$ is the pressure difference between both sides of the tube, and $Q_\mathrm{vol}$ is the volume Poiseuille flow given by
\begin{equation}
    Q_\mathrm{vol} = \frac{ \pi r^4}{8 \mu t}\Delta p
    \;.
    \label{eq:Flow}
\end{equation}
This leads to an approximated formula for the air-flow resistivity
\begin{equation}
\sigma = R_\mathrm{f}\frac{ d^2}{l} = 2 \mu \frac{ d^2}{\pi r^4} \frac{t}{R_\mathrm{p}}
\;,
\label{eq:FlowResistivity}
\end{equation}
as the total thickness of the cell $l= 4R_\mathrm{p}+2t\approx 4R_\mathrm{p}$ when $R_\mathrm{p}\gg t$.
In Eq.~\eqref{eq:FlowResistivity}, the ratio $t/R_\mathrm{p}$ depends only on the pore geometry (foaming process), while $d^2/\pi r^4$ is purely related to the perforations (post-processing).
It is obvious that the flow resistivity increases quickly with decreasing perforation diameter, which becomes the dominant tuning parameter for this transport parameter. However, given the small value of the wall thickness, small variations can have an important influence. The flow resistivity determines the maximum absorption level: if it is too low, there are no viscous and thermal losses; if it is too high, the acoustic wave cannot penetrate the material and the waves are reflected.

The tortuosity can reach very high values compared to fully open-cell foams (i.e, $\tau^\infty \approx 1.05$~\cite{doutres2013semi}). A first reason is the comparatively narrow windows ($r/R_p<1$), which lead to a large local dispersion of the microscopic inviscid velocity field through the material [Eq.~\eqref{eq:lbd}].
The strongest effect however is reached by increasing the perforation distance $d$.
The flow distortion at the entrance of the material corresponds to an increase of the tortuosity~\cite{atalla2007perforated} and leads to a sound absorption peak at lower frequency~\cite{chevillotte2010perforated}. Since a small perforation rate also results in high flow resistivity, this comes at the cost of a reduced absorption level and an optimal configuration should be determined. 

\begin{figure}[h]
\centering
\includegraphics[scale=1]{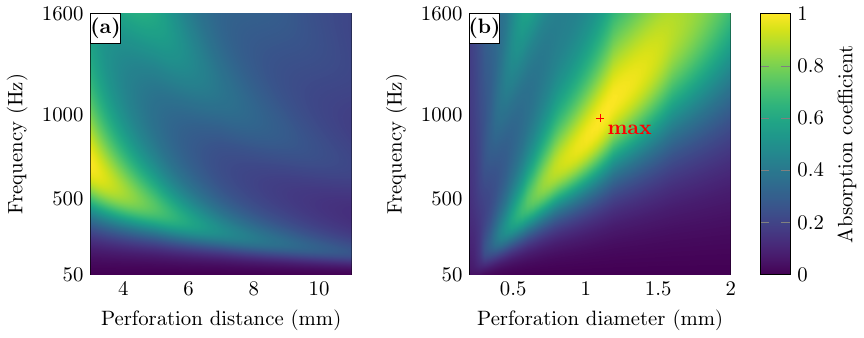}
\caption{(Color online) Normal incidence sound absorption by a 40~mm thick sample as a function of the perforation distance \textbf{(a)} and diameter \textbf{(b)} in a foam with 3~mm pore size and with a wall thickness of $0.1$~mm.}
\label{fig:AbsorptionMap}
\end{figure}

\begin{figure}[h]
\centering
\includegraphics[scale=1]{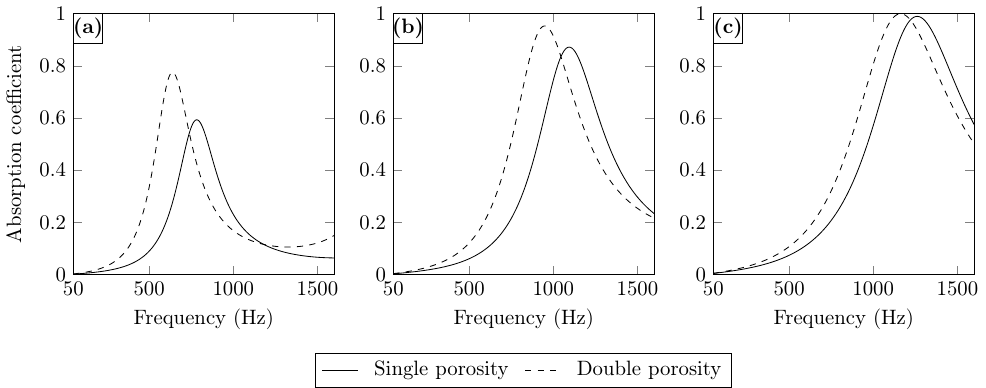}
\caption{Acoustic absorption coefficient for 25~mm thick foams with different pore sizes: 1 mm \textbf{(a)}, 2 mm \textbf{(b)}, and 3 mm \textbf{(c)}. The perforations have a diameter of $2r = 0.5$~mm and are placed with a distance $d=5$~mm. The model with single porosity, omitting the effect of the walls' microporosity, is shown in solid lines
and that accounting for double-porosity is shown in dashed lines.}
\label{fig:Absorption3mat}
\end{figure}

Based on the calculated transport parameters, the effect of the two perforation parameters ($d$, $r$) on the sound absorption of a 40~mm thick foam layer with constant foam morphology parameters ($R_\mathrm{p}=3$~mm and $t=0.1$~mm) are shown in Fig.~\ref{fig:AbsorptionMap}. It can be seen that, for an appropriate choice of perforation parameters, absorption levels higher than $\alpha=0.8$ can be achieved at frequencies as low as 400~Hz.
In this example, this means the peak absorption is reached when the sample thickness is 1/20 of a wavelength in air.
The effect of the pore size is illustrated by three simulated absorption curves for 1, 2, and 3-mm pore diameters. with a perforation pattern defined by $r=0.25$~mm and $d=5$~mm in Fig.~\ref{fig:Absorption3mat}.
A material with smaller pores reaches its peak absorption at lower frequencies, but the absorption is lower due to a too high flow resistivity.

For comparison with the experimental data in the next section, we already point out that the effect of the microporous walls, as described in~\cite{chevillotte2013doublepor}, appears to be favourable for this type of absorbers. The absorption curves for materials with pore sizes 1, 2, and 3~mm were calculated with and without the presence of microporosity, using the macroscopic JCA parameters that can be read from Fig.~\ref{fig:ParametricCellRadius}. 
The JCA parameters of the microporous matrix were taken from~\cite{chevillotte2013doublepor}: $\sigma_m=948000\,\mathrm{kPa.s.m^{-2}}$, $\tau^{\infty}_m = 1.48$, $\Lambda_m=0.83$~\textmu m, and $\Lambda^\prime_m = 1.59 $~\textmu m. The microporosity is however changed to $\phi=0.3$, as determined during the material preparation.
The microporosity results in a slightly decreased overall flow resistivity and increased total porosity which improves the sound absorption levels: the peaks are reaching higher values and appear at lower frequencies (Fig.~\ref{fig:Absorption3mat}, dashed lines).

\section{Experimental study}
\label{sec:Experimental}
A series of experiments was conducted to corroborate the numerical results and predicted absorption curves.
Three types of gypsum foams are available, with a pore size of approximately 1, 2, and 3~mm and a wall thickness of around 0.1 to 0.2~mm.
Optical and electron microscope images of the three foams are shown in Figs.~\ref{fig:ThreeSamples} and \ref{fig:Zoom} respectively.

All three foam types were perforated by 0.5~mm diameter needles, in two patterns with perforation distances $d=5\,\mathrm{mm}$ and  $d=10\,\mathrm{mm}$ respectively. Non-perforated foams were investigated as well, in order to determine the effect of the micro-porosity on flow resistivity and sound absorption.
Circular samples with a thickness of 22~mm and a diameter of 100~mm were cut out of a larger block and their static air-flow resistivity and normal incidence absorption were measured. 

Both tests typically require the samples to be placed inside a test tube.
It is known however that leakage around the edges of rigid foams can lead to erroneous measurement results~\cite{Pilon2004,zielinski2020reproducibility}. 
The samples were therefore cast in a gypsum ring which was clamped against the measurement tubes, rather than fitting them tightly inside the tube.
Good sealing was assured by adding a thin layer of white vaseline, which leaded to repeatable measurement results.
The sample and mounting conditions are shown in Fig.~\ref{fig:FlowRes}(b) and (c).
\subsection{Flow resistivity}
\label{sec:FlowRes}
\begin{figure}[H]
\centering
\includegraphics[scale=1]{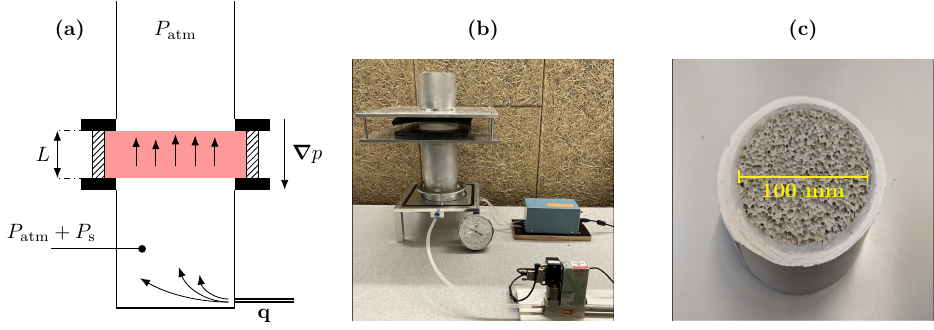}
\caption{(Color online) Sketch \textbf{(a)} and photograph of the test-bench for flow-resistivity measurement \textbf{(b)} and of a sample F3 \textbf{(c).}}
\label{fig:FlowRes}
\end{figure}

The flow resistivity is estimated from Darcy's law using a pressurised chamber, as depicted in the photograph and sketch presented in Fig.~\ref{fig:FlowRes}(a) and (b). From the pressure drop across the sample and the volume flow rate entering the chamber, the static air-flow resistivity is measured according to standard ISO 9053-1:2018.
Each sample is measured from both sides.

\begin{figure}[H]
\centering
\includegraphics[scale=1]{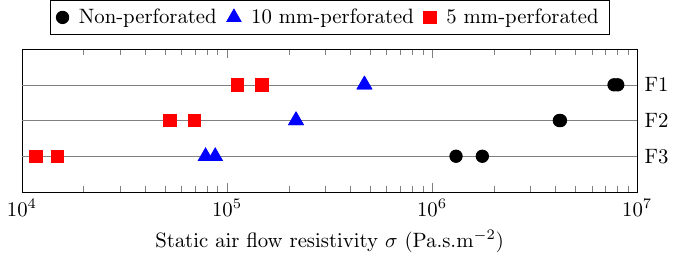}
\caption{(Color online) Results of flow resistivity measurements for different pore sizes (F1: 1~mm, F2: 2~mm, F3: 3~mm) and perforation rates. The two measurement points of each sample represent the values from measurements in two directions.}
\label{fig:ResultFlowRes}
\end{figure}
The non-perforated materials display a high air-flow resistivity, in the order of $10^6-10^7\,\mathrm{Pa.s.m^{-2}}$, as evidenced by the black dots in Fig.~\ref{fig:ResultFlowRes}.
Adding perforations lowers the flow resistivity by roughly a factor 10 ($d=10$~mm) to 100 ($d=5$~mm). 
The transport parameters are calculated according to the homogenization model described in the previous section, according to the geometric properties of the 6 perforated samples.
The experimental results for the perforated foams show good agreement with the values predicted by the models for samples F1 and F2, shown in Table~\ref{tab:TransportParameters}. 
The measured flow resistivities of the perforated sample F3, around 12~$\mathrm{(kPa.s.m^{-2})}$ for $d=5$~mm and 65~$\mathrm{(kPa.s.m^{-2})}$ for $d=10$~mm, is lower than the modeled values. A possible explanation for this is that the thin walls are not perfectly closed, and the cracks formed during the hardening process lower the flow resistance. 

The experiments confirm that materials with smaller pores and larger cell wall thickness have a higher air-flow resistivity [Eq.~\eqref{eq:FlowResistivity}, $\sigma \propto 1/R_p$ ], since more faces are perforated per unit thickness and therefore the flow resistance is increased.
Reducing the perforation distance from 10~mm to 5~mm lowers $\sigma$ by a factor 4, which is in agreement with Eq.~\ref{eq:FlowResistivity}, $\sigma \propto d^2$.
\begin{table}[]
\centering
\caption{Simulated transport parameters for the six perforated foams, assuming all pores are identical Kelvin cells. The perforation diameter is 0.5~mm, the cell wall thickness $t$ is 0.1~mm for the 3~mm pores, and 0.2~mm for the 1 and 2~mm pores.} 
    \begin{tabular}{c|c||c|c|c|c|c}
         pore diameter & perf. distance & $\Lambda^\prime$  & $\Lambda$  & $\phi$ & $\tau^{\infty}$  & $\sigma$  \\
         (mm) & (mm) & (mm) & (mm) & (-) & (-) & $\mathrm{(kPa.s.m^{-2})}$\\
        \hline 
        \hline
          3 & 5 & 0.80 & 0.26 & 0.69 & 7.3 & 33.6\\
          3 & 10 & 0.83 & 0.26 & 0.52 & 22.2 & 123\\
          2 & 5 & 0.53 & 0.25 & 0.49 & 10.9 & 68\\
          2 & 10 & 0.56 & 0.25 & 0.42 & 37.2 & 258\\
          1 & 5 & 0.31 & 0.32 & 0.44 & 15.5 & 114\\
          1 & 10 & 0.29 & 0.26 & 0.33 & 57.2 & 456\\
    \end{tabular} 
    \label{tab:TransportParameters}
\end{table}
\subsection{Acoustic absorption}
\label{sec:Impedance}
\begin{figure}[H]
\centering
\includegraphics[scale=1]{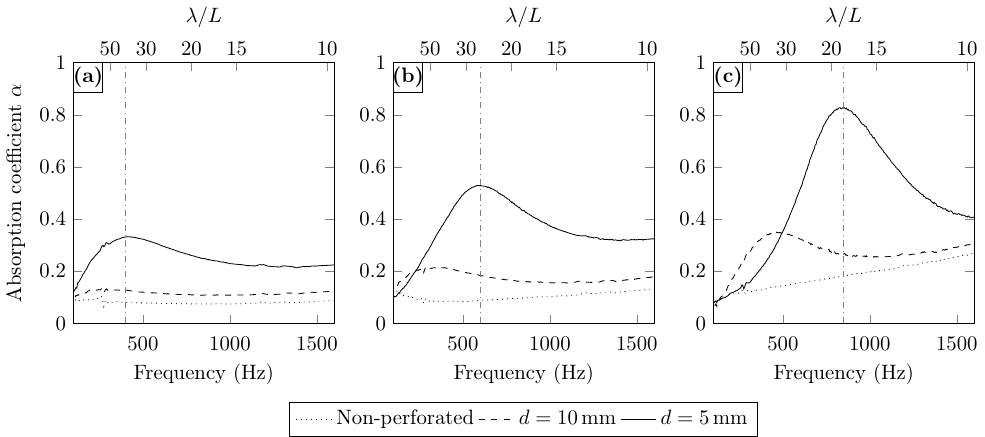}
\caption{Measured acoustic absorption curves at normal incidence for non-perforated and perforated gypsum foam samples. F1 foam sample with pore diameter approximately equal to 1 mm \textbf{(a)}, F2 foam sample with pore diameter approximately equal to 2 mm \textbf{(b)}, and F3 foam sample with pore diameter approximately equal to 3 mm \textbf{(c)}. The upper axis shows the wavelength-sample thickness ratio $\lambda/L$, to emphasize the sub-wavelength absorption peak.
}
\label{fig:absorption}
\end{figure}

The acoustic absorption at normal incidence is measured using an impedance tube Brüel\&Kjaer type 4206, according to ISO 10534-2:1998. 
Since it is known that acoustic absorption of perforated materials can be amplitude-dependent~\cite{tayong2013highsound,zhang2021highamplitude,maa1994microperforated}, the amplitude of the incoming wave is lowered until no change in absorption can be seen, at an SPL of approximately 80~dB.
The results are shown in Fig.~\ref{fig:absorption} for three different base materials (F1-F3) and three perforation cases (non-perforated, $d=10$~mm and $d=5$~mm).

All three non-perforated samples (dotted lines) show low acoustic absorption properties, slightly increasing with frequency.
This points out that the pores are not entirely closed, and the combination of microporosity and cracks yields a certain base absorption.
The F3 sample, which has the largest pores and thinnest walls, clearly reveals the largest initial open porosity among the three samples, although the maximum absorption at 1600~Hz remains below $\alpha=0.3$.

With a perforation pattern using $d=10$~mm the absorption levels are rising, and a pronounced absorption peak is appearing for the F2 and F3 samples, as illustrated in Fig.~\ref{fig:absorption} in dashed lines. The maximum absorption is reached at very low frequencies, typically at $\lambda/40-50$, which confirms the high tortuosity predicted by the models. As expected, the high flow resistivity limits the maximum absorption levels. The effect of the perforations on the sound absorption of the F1 foam is limited, which can be explained by the fact that the perforations are large compared to the pores, and the flow resistivity remains too high.

A perforation pattern with $d=5$~mm (solid lines) results in a measured absorption peak at a frequency corresponding to $\lambda/L\approx17$ for the 3~mm pore foam.
Samples with smaller pores show a predicted tortuosity with larger values, and hence a sound absorption peak appearing at lower frequency.
However, the maximum absorption decreases with decreasing pore size due to a higher flow resistivity.
Overall, opening the cell walls by perforations seems to yield two advantages compared to conventional low-frequency absorbers.
Firstly, for all investigated foam samples the absorption peak width is fairly wide compared to typical resonant acoustic absorbers.
Moreover, the coefficient of absorption beyond the frequency of the sound absorption peak does not drop to very low values, as is typically the case for perforated plates or resonant sound absorbers such as Helmholtz resonators.

\begin{figure}[H]
\centering
\includegraphics[scale=1]{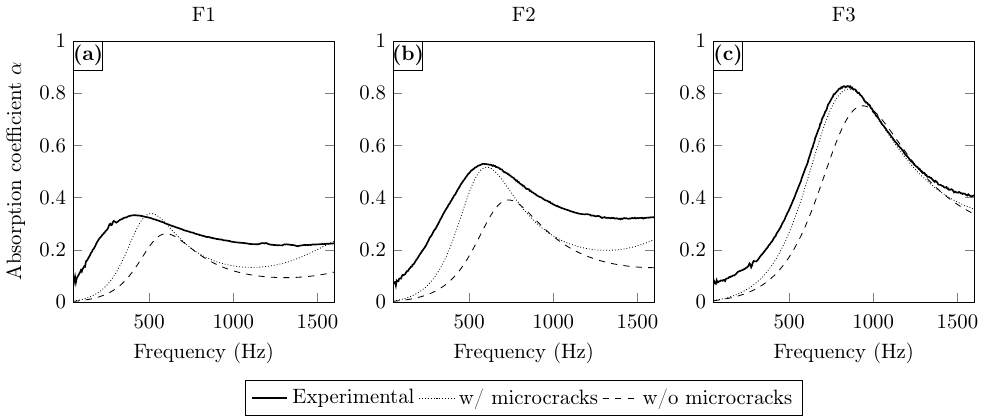}
\caption{Comparison of measured sound absorption (solid line) to the theoretical models with (dashed line)  and without (dotted line) accounting for pre-existing micro-cracks in the solid phase. The viscous characteristic length is reduced from $\Lambda = 0.26$~mm (black dotted line) to $\Lambda=0.08$~mm (red dashed line).}
\label{fig:Validation}
\end{figure}

The measured results clearly show to potential of combining the properties of closed-cell microporous base foams with a well chosen perforation pattern to tune the resulting absorption curve. 
The absorption resulting from a foam-perforation combination is not unique, e.g. illustrated by the fact that F1 with $d=5$~mm yields similar results as F3 with $d=10$~mm. 
This confirms the interest of perforated microporous closed-cell foam structures whose sub-wavelength acoustic absorption can be tailored through the pore morphology defined by the foaming process, and subsequent perforation.
Comparing Fig.~\ref{fig:absorption} and Fig.~\ref{fig:Absorption3mat} reveals a discrepancy between measured and modelled absorption curves, even if the correction for double porosity is taken into account. Of all 5 JCA parameters, only the flow resistivity magnitude prediction was confirmed experimentally. 
The modelled high tortuosity is needed to achieve a low-frequency absorption peak, but the predicted maximal absorption still occurs at higher frequencies than the measured curves.
We assume that those two transport parameters are modelled correctly, and one of the remaining three transport parameters is not well captured by the models. 
A heuristic approach, varying those values is chosen to get insight into which transport parameter  was not accurately captured by the model. 
The thermal characteristic length does not significantly change the absorption curves when it is changed over one order of magnitude.
Varying the macroporosity does have an effect of the achieved maximum absorption, but does not change the peak frequency.
The viscous characteristic length, however, can be tuned to achieve good agreement between the measured and modelled curves as shown in Fig.~\ref{fig:Validation} for material F3 with $d=5$~mm. The value has to be lowered to $\Lambda = 80~\mu$m, which is significantly lower than the initial modelled value ($\Lambda=260\mu$m, Tab.~\ref{tab:TransportParameters}).
The viscous characteristic length is typically assumed to be similar to the window radius in open pore materials.
Since we know that small cracks are present in the cell walls, they can lower the value of this transport parameter (Sec.~IV B of Ref.~\cite{chevillotte2013doublepor}).
Although a full analysis of the required crack density to achieve the corrected value of $\Lambda$ is outside the scope of this work, an outline of possible numerical investigation is shown in \ref{sec:Windows}.
Models confirm that the presence of one ore more smaller windows additional to the main perforation lowers the viscous characteristic length.
\section{Conclusions}
\label{sec:Conclusions}
In this work, we investigated the potential of perforating rigid mineral foams with high porosity and low mass density to achieve sub-wavelength sound absorption under normal incidence.
Several features of the proposed material result in remarkable acoustic properties.
First of all, perforating the otherwise closed-cell material gives rise to very high tortuosity, yielding high acoustic absorption at low frequencies. Second, an appropriate choice of the perforation pattern for a given pore geometry defines a combination of open porosity, tortuosity, and flow resistivity to achieve sub-wavelength sound absorption. Third, the micro-porosity of the cell walls even further lowers the frequency of the maximum absorption peak, and increases the peak absorption level. 

The absorption spectrum can be tuned to desired values by two very different processes. First, the chemical process to form the foams is remarkably reproducible and adaptable, so that the porosity, pore size, wall thickness, and even microporosity can be changed within fairly large bounds. Second, the perforation process yields the acoustic absorption by optimizing the perforation distance and diameter. 
A predictive, numerical homogenization method can be used to determine a suitable combination of design parameters.

The combination of these factors leads to a lightweight, non-flammable material showing high acoustic absorption for layers as thin as 1/20 of the desired wavelength. However, as with all perforated solutions, the absorption in the higher frequency range remains quite limited. Since the main physical mechanism does not come from a mechanical or acoustic resonance phenomenon, the absorption level however never drops to very low values after the first peak, and the material is still reasonably effective. Because the material can be produced at a large scale (mass production), we believe it has a wide field of applications, both indoor and outdoor, in spaces with high low-frequency noise where it could present a space-saving economical solution.

On a more fundamental level, the origins of the high tortuosity achieved within the porous material were investigated in detail. The first reason was related to the dispersion of microscopic velocities due to an abrupt cross-sectional change through the opened cell walls, effectively increasing the path length inside the material. This, alone, however, did not explain the change in tortuosity when the perforation distance was modified. By creating a model which includes the inlet to the material explicitly, the recirculation of the fluid due to the presence of closed and perforated pores on the material's surface was accounted for. Indeed, the inviscid flow, at the high-frequency asymptotic behaviour, follows this complex path, thereby tremendously increasing the tortuosity. This can be seen as an important practical advantage when compared to adding a smooth perforated screen to an open-cell material.

A double-porosity model combining  perforated Kelvin cells with a microporous matrix showed good agreement with the measurements and captured most of the physics. However, the viscous characteristic length was found to be smaller than the value predicted by the model. Additionally, the measured absorption outside the main peak is consistently higher than predicted. Both deviations between the measurements and models are most likely a result of wall openings, smaller than the perforation radius, due to the foaming process. The roughness and asperities of the macroscopic pores might also affect the results, in particular lower the viscous characteristic length, as was shown by \cite{cortis2003influence}. Therefore, an improvement of the used multi-scale approach would consist in modeling another scale coupling the microporous matrix and the pore size level by introducing morphological information such as cracks and surface roughness. Finally, the complex geometry with the inherent spreading of pore sizes and wall thicknesses leads to a richer behaviour than the idealized model.

\section*{Acknowledgments}
The authors gratefully acknowledge U.~Gonzenbach, P.~Sabet, and P.~Sturzenegger from de Cavis AG.
This work was jointly funded by Innosuisse project 56633.1 and de Cavis AG.
We acknowledge the support of the COST Action DENORMS CA 15125  funded by the European Cooperation in Science and Technology through the funding of a short therm scientific mission (STSM). 
\newpage
\appendix
\section{Multiple-scale homogenization theory}
\label{sec:Homogenization}
Mineral foams are two-phase media, consisting of a solid matrix and air saturating the pores.
In the present case, the foams are not only porous at the macroscopic level, but the rigid skeleton itself displays some porosity, and is thus not impervious. The modelling of acoustic propagation in such heterogeneous materials has been described extensively in the past~\cite{johnson1987theory,champoux1991dynamic,Allard2009}. In this section, we briefly outline the pore-scale models leading to the homogenized transport parameters needed for the JCA model calculations. This will allow us to perform the parameter study in the next section.

Assuming the foams' pore size distribution is macroscopically constant in space, and that the heterogeneities are very small compared to the acoustic wavelength, the foam can be considered to have homogenized properties similar to those of a fluid. The wave propagation, in particular dispersion and attenuation phenomena, are encapsulated in complex valued and frequency dependent properties, that describe the acoustic field in the foam. These apparent properties can either be estimated from inversion procedures on experimental data, or estimated by computing the local field of flow velocity in the fluid phase of the material.

While the velocity field can be computed for all frequencies, we use a well established asymptotic approach to drastically reduce the required computational resources. The JCA model relies on five asymptotic transport parameters, which are sufficient to estimate the frequency-dependent bulk modulus $B$ and mass density $\rho$. The transport properties can be calculated from first principles using the representative volume element (RVE) of the porous material, defined by a domain $\Omega$. To do this, two numerical models are necessary: one for viscous flow and one for inviscid flow through the fluid domain $\Omega_f$ within the rigid-wall structure defined by the surfaces $\Gamma$. For complex geometries, these models are performed numerically.
Most published studies use finite element simulations \cite{zielinski_benchmarks_2020,terroir_general_2019,boulvert_acoustic_2020,cavalieri_acoustic_2019,cavalieri_graded_2020}, whereas in this work we use the finite volume approach of commercial computational fluid dynamics (CFD) software (Ansys Fluent). Although the discretization and solution methods are very similar, the use of CFD software typically requires specific boundary conditions such as pressure and potential differences, rather than the mathematical body forces usually applied in finite element calculations. The calculation of the five parameters has been described in many publications, many of which have slightly different conventions and notations. We follow the benchmark description in~\cite{zielinski_benchmarks_2020}, which was also validated for materials with low porosity and high tortuosity.

Two parameters are purely geometrical, and do not require a physical simulation: the open porosity $\phi$ and the thermal characteristic length $\Lambda^\prime$ or generalized hydraulic radius. They are defined as
\begin{equation}
\left\{
\begin{array}{lll}
\phi &=& \vert \Omega_\mathrm{f} \vert / \vert \Omega \vert \;,\\ \\
\Lambda^\prime &=& \displaystyle 2 \left(\int_{\of} \! \dd V\right)\left(\int_{\Gamma}\dd S\right)^{-1}
\;.
\end{array}
\right.
\end{equation}
The static flow resistivity can be retrieved from a low Reynold's number viscous flow simulation, applying a pressure difference $\Delta p$ between the inlet and outlet faces of the RVE. Solving the Stokes equation 
\begin{equation}
\mu \nabla^2 \vect{v} = \vect{\nabla} p\;,\vect{\nabla}\cdot\vect{v}=0
\end{equation}
yields a velocity flow field $\vect{v}$ within the fluid domain, which is constrained by no-slip conditions on $\Gamma$. The constant $\mu$ is the dynamic viscosity of the fluid. The static flow resistivity $\sigma_j$ in direction $j$ is then defined by 
\begin{equation}
\sigma_j = \frac{\Delta p}{t_\mathrm{RVE} \left \langle v_j \right \rangle}
\;,
\end{equation}
where $t_\mathrm{RVE}$ is the thickness of the RVE in direction $j$. The volume-average operator is defined over the total RVE volume $\vert \Omega \vert$ by $\displaystyle \left\langle \bullet \right\rangle = \frac{1}{\vert \Omega \vert} \int_{\Omega} \!\bullet  \,\dd V = \frac{1}{\vert \Omega \vert} \int_{\of} \!\bullet  \,\dd V$. 

The two high-frequency JCA parameters in direction $j$ are the tortuosity $\tau^{\infty}_j$ and the viscous characteristic length $\Lambda_j$.
At the high-frequency limit, the viscosity can be neglected and the calculation is typically carried out with an analogue electrical conduction boundary value problem following the Laplace equations for the potential flow:
\begin{equation}
\begin{array}{lll}
   \vect{E}   =  - \vect{\nabla}q, \vect{\nabla}\cdot\vect{E}  =  0  &\text{  in  }& \Omega_f\\
    \vect{E}\cdot\vect{n} = 0 &\text{  on  }& \Gamma
\end{array}
\end{equation}
The driving force is a potential difference $q$ over the inlet and outlet of the RVE
, and the resulting electrical field $\vect{E}$ 
are constrained by perfectly isolating domain walls.
The two high-frequency transport parameters are then provided by the equations
\begin{equation}
\left\{
\begin{array}{lll}
\displaystyle \tau^{\infty}_j &=& \displaystyle\frac{\left \langle \vect{E}_j\cdot\vect{E}_j \right \rangle_f}{\left \langle \vect{E}_j \right \rangle_f^2}
\;,\\ \\
\displaystyle \Lambda_j &=&  
2\frac{\displaystyle\int_{\of} \vect{E}_j\cdot\vect{E}_j\, \dd V }
{\displaystyle \int_{\Gamma} \vect{E}_j\cdot\vect{E}_j\, \dd S}
\;.
\end{array}
\right.
\label{eq:lbd}
\end{equation}
The volume-average operator for the tortuosity is defined over the fluid volume by $\displaystyle \left\langle \bullet \right\rangle_\mathrm{f} = \frac{1}{\vert \of \vert} \int_{\of} \!\bullet \,\dd V = \displaystyle \left\langle \bullet \right\rangle/\phi$.
Integrating the inlet layer into the model explicitly includes the low perforation rate into the tortuosity calculation.
In previous work, it was shown that the tortuosity has to be corrected in layered combinations of perforated panels and porous absorbers to address the flow distortion at the entrance of the absorber~\cite{atalla2007perforated}. The low perforation rate significantly increases the perceived tortuosity of the absorbing system, which can be introduced via a correction factor of the bulk properties. An inlet layer in the model allows for the explicit modelling of the flow lines at the material entrance, and thus yields a value which does not have to be corrected. Details about the influence of the inlet layer are given in~\ref{sec:inlet}.
The effective complex and frequency-dependent bulk modulus $B(\omega)$ and mass density $\rho(\omega)$ are then readily obtained from the transport parameters:
\begin{equation}
\left\{
\begin{array}{l}
\displaystyle B(\omega) = \displaystyle \gamma P_0 \left( \gamma - \frac{\gamma - 1}{\displaystyle 1 + \frac{8 \nu' }{\ii \omega \Lambda'^2}G'(\omega)} \right)^{-1}
\; ,\\
\\
\displaystyle \rho(\omega) = \displaystyle \rho_0 \left( \tau^{\infty} + \frac{\nu \phi}{\ii \omega K_0} G(\omega) \right)\;,
\end{array}
\right.
\end{equation}
where $\ii$ is the imaginary number, $P_0$ is the ambient pressure, $\rho_0$ the equilibrium mass density of the fluid, $\gamma$ the adiabatic constant, $\nu = \mu/\rho_0$ the fluid's kinematic viscosity, and $\nu^\prime=\nu/\mathrm{Pr}$ with $\mathrm{Pr}$ the fluid's Prandtl number.
The visco-static permeability is defined as $K_0 = \mu/\sigma$.
The functions $G(\omega)$ and $G'(\omega)$ are given by
\begin{equation}
    \left\{
        \begin{array}{l}
             \displaystyle G'(\omega) = \left( 1 + \left( \frac{\Lambda'^2}{16} \frac{\ii \omega}{\nu'}\right) \right)^{1/2} \; ,\\
             \displaystyle G(\omega) = \left( 1 + \left( \frac{2 \tau^{\infty} K_0}{\phi \Lambda} \frac{\ii \omega}{\nu}\right) \right)^{1/2} \; .
        \end{array}
    \right.
\end{equation}
The effect of the double porosity has been investigated in~\cite{olny2003double}, and was applied in~\cite{chevillotte2013doublepor} for the case of gypsum foams.
If the transport parameters of both the micro- and macro-pores are known, their homogenized properties can be combined into a single effective medium:
\begin{equation}
    \left\{
        \begin{array}l
             \displaystyle B_\mathrm{db}(\omega) = \left( \frac{1-\phi_\mathrm{macro}}{B_\mathrm{micro}} + \frac{1}{B_\mathrm{macro}} \right)^{-1}\;,\\
             \\
             \displaystyle \rho_\mathrm{db}(\omega) = \left( \frac{1-\phi_\mathrm{macro}}{\rho_\mathrm{micro}} + \frac{1}{\rho_\mathrm{macro}} \right)^{-1}\;.
        \end{array}
    \right.
\end{equation}
The transport parameters of the microporous phase were taken from~\cite{chevillotte2013doublepor}, except the value of the microporosity which depends on the manufacturing process. The values used for the microporous phase are $\sigma_m = 948$~MPa.sm$^{-2}$, $\Lambda_m = 0.83$~\textmu m, $\Lambda'_m = 1.59$~\textmu m, $\tau^{\infty}_m=1.48$, and $\phi_m=0.30$.

\section{Influence of the inlet layer on the calculated transport parameters}
\label{sec:inlet}
Since the calculation of transport parameters using FEM is typically done applying periodic boundary conditions, the explicit presence of an inlet layer in the model is somewhat unconventional. However, in the presented case of perforated highly porous materials, the high tortuosity which is responsible for sub-wavelength sound absorption can only be explained by the flow distortion at the material inlet. In this work, this effect is modeled by adding a slice of fluid to the inlet of the periodic microstructure. In this short section, we show the effect of the presence of an inlet zone on the modelled transport parameters.

\begin{figure}[h]
\centering
    \begin{subfigure}{0.3\textwidth}
        \includegraphics[width=\textwidth]{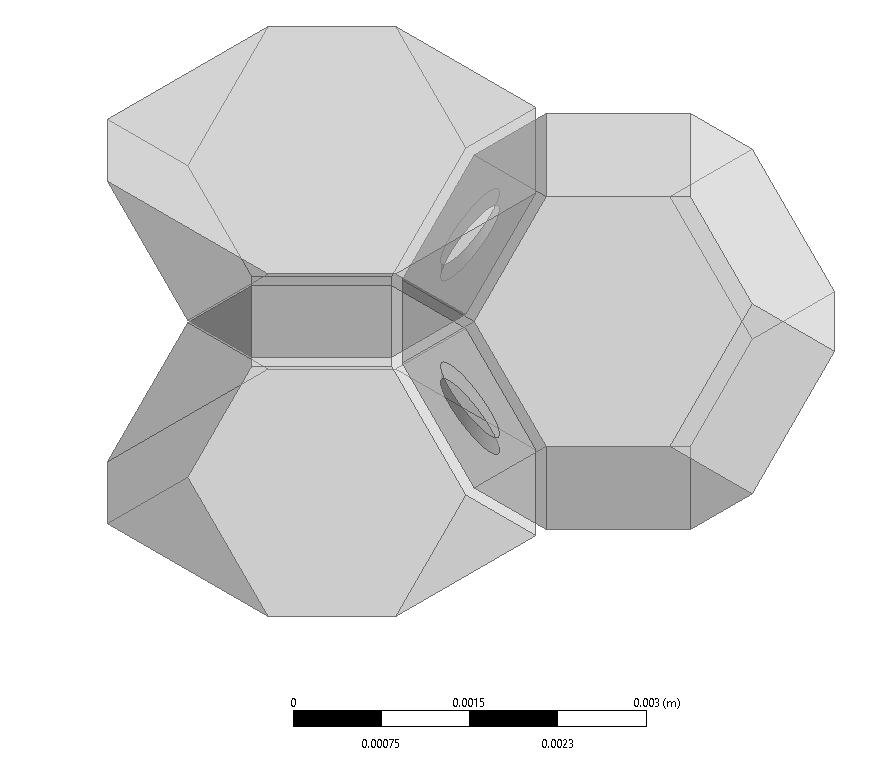}
    \end{subfigure}
    \hfill
    \begin{subfigure}{0.3\textwidth}
        \includegraphics[width=\textwidth]{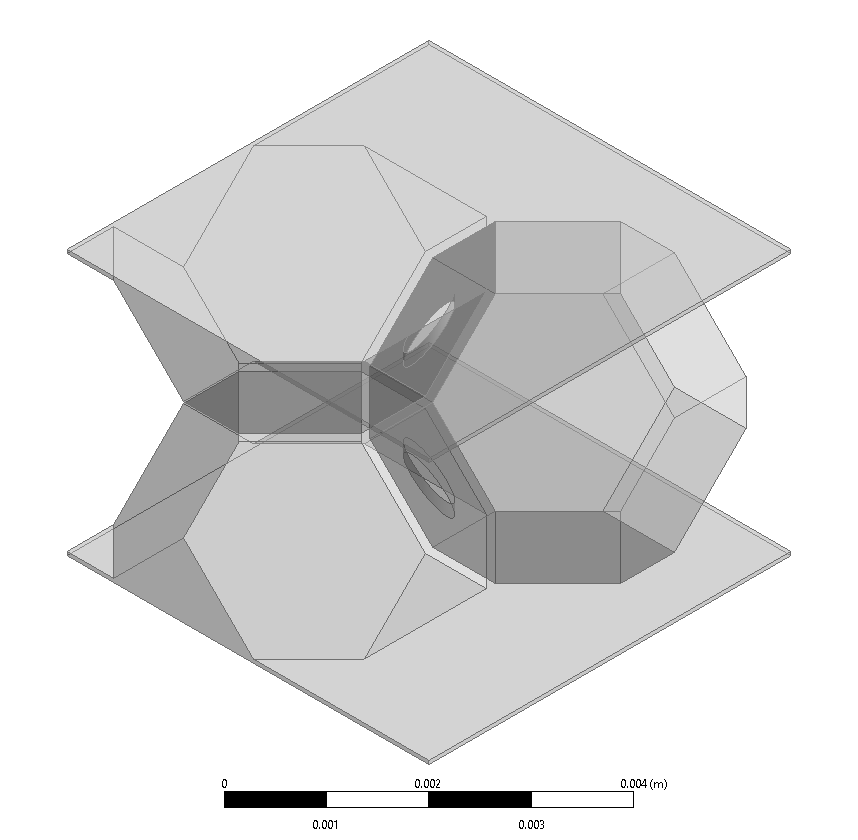}
    \end{subfigure}
    \hfill
    \begin{subfigure}{0.3\textwidth}
        \includegraphics[width=\textwidth]{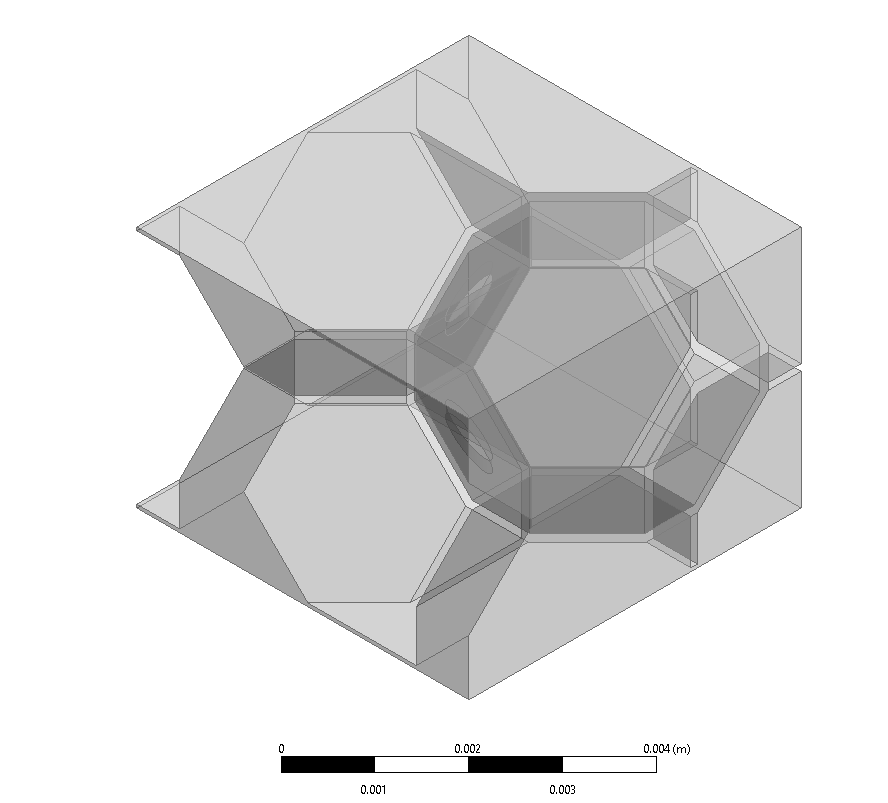}
    \end{subfigure}
\caption{Three geometries to demonstrate the inclusion of a fluid inlet to the perforated pores. (I, left panel) No inlet: The fluid directly penetrates inside the perforated pores. (II, central panel) Inlet not connecting the closed pores: The fluid penetrates inside the perforated pores through an intermediate slice of fluid. (III, right panel) Inlet connecting the closed pores: The fluid penetrates inside all the pores in contact with the intermediate slice of fluid and can recirculate inside the isolated pores towards the perforated pores via the intermediate slice of fluid.}
\label{fig:inlet}
\end{figure}

Three scenarios are investigated, shown in Fig.~\ref{fig:inlet}. First, we assumed the perforated Kelvin cells without a slice of  fluid, which agrees with the case of the bulk material.
In this scenario, no distortion of the inlet velocity field due to the large spacing between perforations is taken into account.
The second scenario adds on a fluid slide to the connected Kelvin cells, but leaves the closed pores out.
This is similar to adding a perforated plate to the bulk material, and brings some additional flow distortion at the entrance of the material. The third scenario, which is further used in this paper, also includes an air inlet at the exposed surface of the non-perforated closed pores which can thus affect the flow path.
The high-frequency, inviscid, flow can therefore recirculate from a closed to a perforated pore, thereby further increasing the path followed by the air particles and the tortuosity.
\begin{table}[]
    \centering
    \caption{Simulated transport parameters for perforated Kelvin cells representing material F3, with and without an air inlet layer with varying thickness. Scenario I refers to the absence of an air layer, scenario II to a smooth air layer, and scenario III to an air layer connecting the non-perforated pores.}
    \begin{tabular}{c|c|c|c|c|c|c|c}
        Scenario & inlet layer & perf. distance & $\Lambda'$  & $\Lambda$  & $\phi$ & $\tau^{\infty}$  & $\sigma$  \\
         & (mm) & (mm) & (mm) & (mm) & (-) & (-) & $\mathrm{(kPa.s.m^{-2})}$\\
        \hline
         I & 0 & 5 & 1.03 & 0.26 & 0.46 & 4.4 & 15.0\\
         II & 0.1 & 5 & 0.78 & 0.26 & 0.47 & 4.44 & 15.0\\
         II & 0.5 & 5 & 0.89 & 0.27 & 0.52 & 4.70 & 15.2\\
         II & 0.1 & 10 & 0.39 & 0.26 & 0.14 & 5.28 & 17.9\\
         II & 0.5 & 10 & 0.66 & 0.27 & 0.22 & 8.05 & 26.2\\
         III & 0.1 & 5 & 0.80 & 0.26 & 0.69 & 6.53 & 15.2\\
         III & 0.5 & 5 & 0.88 & 0.27 & 0.72 & 6.46 & 15.3\\
         III & 0.1 & 10 & 0.83 & 0.26 & 0.52 & 19.7 & 67.0\\
         III & 0.5 & 10 & 0.98 & 0.26 & 0.57 & 20.2 & 66.4\\
    \end{tabular} 
    \label{tab:inlet}
\end{table}
For the standard case investigated in the parametric study, with a pore diameter of 3~mm and wall thickness of 0.1~mm, a perforation diameter equal to 0.5~mm and a lattice constant $d=5$~mm, adding a smooth inlet (scenario II) does not considerably increase the tortuosity or flow resistivity~\ref{tab:inlet}. There is very little change when the inlet layer's thickness is increased. By contrast, with a perforation spacing of 10~mm, the tortuosity and flow resistivity clearly increase. However, the thickness of the inlet layer has a considerable effect on the calculated values, and it might therefore not be suitable approach to the modelling strategy (the output transport parameters are too sensitive to the input fluid layer thickness).

By including the closed pores exposed to the inlet surface (scenario III), an important change of the calculated values can be noticed. The tortuosity value increases from about 4.4 to 6.5 with a 5~mm perforation spacing and even to approximately 20 for a 10~mm spacing. This is in line with the experimental results, where the frequency at which the peak of sound absorption occurs decreases with increasing perforation spacing, which is consistent with an increase of tortuosity. The flow resistivity increases from 15 to 67 $\mathrm{kPa.s.m^{-2}}$ when the spacing increases from 5 to 10~mm. This result is in agreement with the simplified calculation presented throughout Eq.~\eqref{eq:FlowResistivity}. It must be noticed that the thickness of the inlet layer in scenario III does not significantly change the simulated transport parameters. We can therefore conclude that this approach is compatible with the measurements, and captures the important physical properties of the absorption mechanism: the tunability of tortuosity and flow resistivity by a well-informed choice of the perforation pattern for a given foam geometry.
\section{Influence of small windows on the viscous characteristic length}
\label{sec:Windows}
The only transport parameter which is not accurately captured by the numerical model is the viscous characteristic length $\Lambda$ (Eq.~\eqref{eq:lbd}). It is typically assumed that this value is in close agreement with the window radius in open pore materials (in this case the perforation radius, 0.25~mm), which is confirmed by the numerical model (Tab.~\ref{tab:TransportParameters}). However, for the given tortuosity and flow resistivity, it is found that the viscous characteristic length has to be reduced to capture the measured absorption curves. Unfortunately, a direct measurement of the viscous characteristic length is not straightforward and requires ultrasonic experiments in different gases~\cite{leclaire1996charlengths}.
\begin{figure}[h]
    \centering
    \begin{subfigure}{0.4\textwidth}
        \includegraphics[width=\textwidth]{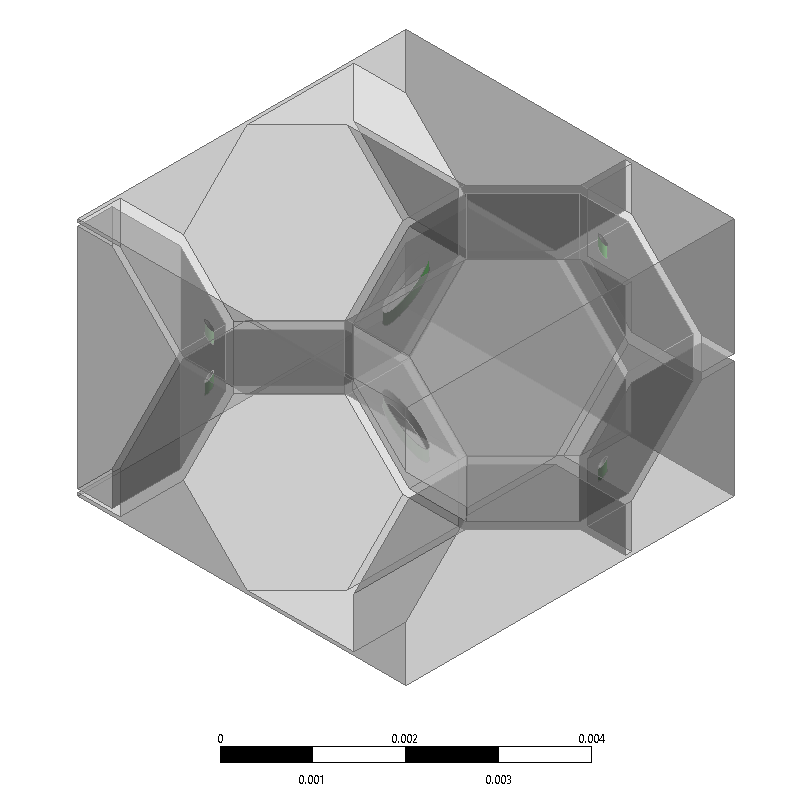}
    \end{subfigure}
    \quad
    \begin{subfigure}{0.4\textwidth}
        \includegraphics[width=\textwidth]{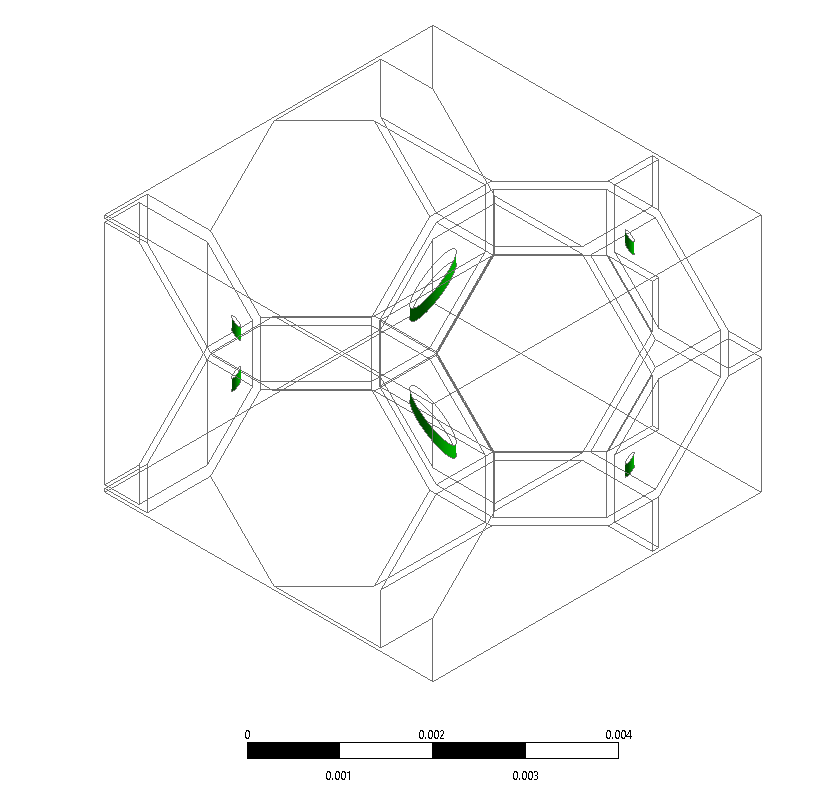}
    \end{subfigure}
    \caption{Simulation of two microcracks left and right of the main perforation. The cylindrical apertures are 5 times narrower than the perforation, as can be clearly seen from the line representation in the right panel.}
    \label{fig:Cracks}
\end{figure}
\begin{table}[]
    \centering
    \caption{Simulated transport parameters for perforated Kelvin cells with an additional cylindrical microcrack.}
    \begin{tabular}{c|c|c|c|c|c|c}
         crack radius & perf. radius & $\Lambda'$  & $\Lambda$  & $\phi$ & $\tau^{\infty}$  & $\sigma$  \\
         (mm) & (mm) & (mm) & (mm) & (-) & (-) & $\mathrm{(kPa.s.m^{-2})}$\\
        \hline
          0.05 & 0.25 & 0.76 & 0.18 & 0.73 & 6.76 & 23.3\\
          0.033 & 0.25 & 0.76 & 0.19 & 0.73 & 6.85 & 23.3\\
          0.025 & 0.25 & 0.76 & 0.20 & 0.73 & 6.88 & 23.4\\
          0 (no crack) & 0.25 & 0.80 & 0.26 & 0.69 & 6.53 & 22.1\\
    \end{tabular} 
    \label{tab:Crack}
\end{table}
In the models of the parametric study, the walls are assumed to be impervious. The microscope images in Fig.~\ref{fig:ThreeSamples} show that in the thin walls of these foams several cracks occur, which are much smaller than the perforations however 
.
We mimic this by adding a single additional perforation to the RVE, with a radius much smaller than the main perforation as shown in Fig.~\ref{fig:Cracks}.
The results are displayed in Tab.~\ref{tab:Crack}. Even for a single microcrack in the order of a few tens of micrometers the viscous characteristic length is reduced, depending on its size, up to 30\%.
The models show that adding a second crack will further reduce the viscous characteristic length $\Lambda$ from 0.18 to 0.15~mm. 

We conclude that a network of microcracks effectively decreases the viscous characteristic length, while practically not affecting the other transport parameters. Cracks induced by the perforation process or other imperfections of the solid matrix might further reduce this parameter, to finally achieve a low enough value consistent with the measured absorption curves.

\end{document}